\documentclass[mathleft,fleqn,%
]{an}
\usepackage{graphicx}
\usepackage[varg]{txfonts}
\begin{document}

\Pagespan{1}{}
\Yearpublication{2014}%
\Yearsubmission{2014}%
\Month{0}%
\Volume{999}%
\Issue{0}%
\DOI{asna.201400000}%

\title{Electron acceleration and small scale coherent structures formation by an Alfv\'en wave propagating in coronal interplume region}

\author{K. Daiffallah\inst{1}\fnmsep\thanks{Corresponding author:
        {k.daiffallah@craag.dz}}
\and  F. Mottez\inst{2}
}
\titlerunning{Acceleration in interplume region by Alfv\'en Waves}
\authorrunning{K. Daiffallah \& F. Mottez}
\institute{
Observatory of Algiers, CRAAG, Route de l'Observatoire, BP 63, Bouzar\'eah 16340, Algiers, Algeria
\and 
Laboratoire Univers et Th\'eorie, CNRS, Observatoire de Paris, Universit\'e Paris Diderot, PSL Research university, Meudon, France
}

\received{XXXX}
\accepted{XXXX}
\publonline{XXXX}


\abstract{%
 We use 2.5-D electromagnetic particle-in-cell simulation code to investigate the
     acceleration of electrons in solar coronal holes through the interaction of Alfv\'en waves with an interplume region. The interplume is modeled by 
      a cavity density gradients that are perpendicular to the
     background magnetic field. The aim is to contribute to explain the observation of suprathermal electrons under relatively quiet sun. 
Simulations show that Alfv\'en waves in interaction with the interplume region gives rise to a strong local electric
     field that accelerates electrons in the direction parallel to the background magnetic field. Suprathermal electron beams and small-scale coherent structures are observed within interplume of strong density gradients. These features result from non linear evolution of the electron beam plasma instability.}

\maketitle

\section{Introduction}

The origin of solar wind suprathermal electrons remains an unsolved problem in solar physics. Different acceleration processes based on magnetic reconnection have been proposed. However, observations indicate that electrons are accelerated even under quiet solar conditions (Lin, 1980). Pierrard et al. (1999) found from the observational data of the {\it{WIND}} spacecraft at 1 A.U that suprathermal particles can be originated from low altitude corona.

In this work, we look at a process which takes place in a small
area in the solar corona where  parallel electric fields 
to the background magnetic field can accelerate electrons efficiently. The origin
of these parallel electric fields can be associated with regions of depletion of
density, i.e., plasma cavities

The possible link of the acceleration regions with plasma cavities can be seen for instance in the Earth auroral zone.  Density cavities with a small width (a few km) were observed in this region by space probes like {\it{Viking, Freja}} and {\it{FAST}} at altitudes
between 2000 and 12.000 km (Hilgers et al. 1992; Louarn et al. 1994; Carlson
et al. 1998). 
In the Earth auroral zones, where in-situ measurement were done, 
the plasma cavities are systematically related to the presence of electron beams and strong electrostatic fields. 
It was shown that the propagation of Alfv\'en waves along an inhomogeneous
plasma in Earth magnetosphere with perpendicular density gradients can
efficiently accelerate electrons to a few keV (G\'enot, Mottez \& Louarn 2001;
G\'enot, Louarn \& Mottez 2004; Mottez \& G\'enot 2011).

In this study, we expect another source of suprathermal electrons different from  solar flares, namely the interplume region in the polar coronal holes, where the solar wind escapes with a high-speed.

Plumes are prominent bright and narrow structures of the solar corona, observable both in visible and ultraviolet (UV).  They are associated to open magnetic field lines which expand from the chromospheres to the upper corona. Interplumes appear as darker and sharper regions surrounding the bright plumes (Wilhelm 2006; Wilhelm et al. 2011; Poletto et al. 2015).

Observations show that in the low corona, the density in plumes can be four to seven times more than in interplume regions. However, with increasing height, this density decreases faster than those of interplumes and the density ratio becomes smaller (see Figure 17 in Wilhelm et al. 2011). The electron densities in plume is estimated to be up to about $10^9$ cm$^{-3}$ for $R>R_{\sun}$ (Wilhelm et al. 2011).  The observed density of a typical plume with cylindrical symmetry has a radial gradient profile which declines with the distance from the plume axis (Newkirk \&  Harvey 1968). 
Given that, the density in the interplume region will show a cavity-like profile.

Using the Solar Optical Telescope aboard {\it{Hinode}}, Tsuenta et al. (2008) found that the polar region shows a many vertically oriented magnetic flux tubes with field strengths as strong as 1 kG. These open magnetic fields are dominated by a single polarity (Ito et al. 2010). Tsuenta et al. (2008) estimate the average total magnetic flux for the polar regions situated at latitudes above 70 $^{\circ}$. They found values between 3.1 and 13.9 G. These numbers should be regarded as minimum values. McIntosh et al. (2011) show that the observed densities and phase speeds associated to transverse oscillations in the coronal holes and quiet Sun regions are compatible with the presence of magnetic fields of the order of 10 G.

In general, the electron temperature in coronal holes is estimated to be between 0.7 and 2 MK at height $h$ = 0 (see Figure 18 in Wilhelm et al. 2011).

Waves are present both in plumes and interplumes. Indirect observations of transverse Alfv\'en waves in coronal holes have been reported by many authors (see review by Banerjee et al. 2011).  Using comparison between ({\it{SDO}}/AIA) data of the observed Alfv\'enic motions and synthetic data from Monte Carlo simulation, McIntosh et al. (2011) demonstrate that the wave energy flux is sufficient to accelerate the fast solar wind and heat the quiet corona. Transverse Alfv\'en waves have been detected directly for the first time in solar polar plumes by Thurgood et al. (2014) using observations from ({\it{SDO}}/AIA) instrument. However, their results indicate that Alfv\'en waves cannot be the dominant energy source for fast solar wind acceleration in coronal holes.

Wilhelm et al. (1998) measured velocities in coronal-holes plasma with the SUMER instrument on {\it{SOHO}}. They found strong accelerations of ions in interplume lanes in comparison to those observed in plume regions.  
Observations with {\it{HINODE}}/EIS and {\it{SOHO}}/SUMER by Gupta et al. (2010) show that interplumes support either Alfvénic or fast magneto-acoustic waves, while plumes support more the slow magneto-acoustic waves. Moreover, the authors observed for the first time a signature of acceleration of the waves in interplumes regions.
These observations provide strong indications that the propagation of Alfv\'en waves in interplume regions play a major role on the acceleration of the fast solar wind. This conclusion is in agreement with several earlier works (see Poletto et al. 2015).
Alfv\'en waves propagate from lower solar atmosphere transporting energy to the corona along regions of open magnetic field, such as polar coronal holes. A fraction of this mechanical energy is converted to thermal energy which dissipates and heats coronal holes and accelerates the fast solar wind. Various MHD models based on turbulences and interaction of Alfv\'en waves with inhomogeneous plasmas were proposed to explain the dissipation mechanisms (see review by Ofman 2005).

Moreover,  the non-linear turbulent evolution of the Alfv\'en wave activity 
can favour the transfer of energy from long wavelengths to shorter ones. 
We can therefore expect Alfv\'en waves in the MHD range of long wavelengths, as well as 
of shorter lengths whose frequencies scale as the ion gyro-frequency
(Bale et al. 2005; Podesta \& TenBarge 2012). 
This phenomenon is evidenced with in-situ measurements in the solar wind
(Goldstein \& Roberts 1999).  Actually, shorter wavelengths have even been seen in
space plasma, cascading down to electron scale lengths
(Sahraoui et al. 2013). These observations have been supported by EMHD
numerical simulations (Biskamp et al. 1999), were the Alfv\'enic fluctuations can reach frequencies of the order of the ion gyrofrequency. 
 These theoretical works show that the spectrum of Alfv\'en waves extended
 over a broad interval of frequencies/wavelengths is relevant for a large
 domain of plasma parameters and it can be expected in many areas of the solar
 environment, including the low beta regions of its corona (Zhao, Wu \& Lu 2013).

This type of cascade gives rise to kinetic Alfv\'en waves when the perpendicular wavenumber $k_{\perp} \ne 0$.
Beyond the first several solar radii of height from the solar surface, the plasma in coronal holes becomes collisionless, which makes it possible to use multi-fluid kinetic processes to treat the plasma. Then wave-particle kinetic interactions will act as effective collisions in this plasma, and may lead to heating, as discussed by several kinetic models (see review by Cranmer 2009).

In this context,  Wu \& Fang (2003) investigated the dissipation of a kinetic Alfv\'en waves through an empirical model of a plume and uniformly low-$\beta$ magnetized coronal hole surrounding media. The dense plume is characterized by a radial steady flow with transverse pressure balance. The authors conclude that the dissipation of the kinetic Alfv\'en waves energy provide a sufficient  local electron heating to balance the enhanced radiative losses of the bright plumes.

\begin{table*}
\center
\caption{Simulations parameters in normalised units.}
\label{T-simple}
\begin{tabular}{ccccclcc}     
\hline                   
Run & Particles & Cavity     & $B_0$ & wavelength & $\delta B/B_0$  & $V_{Te}$ & $(m_i/m_e) \beta$\\
    &  per cell & depth $p$  &       &             &                 &          &                 \\
\hline
A & 100  	& No cavity & 0.315 	& 71.7 & 0.3 &  0.07 	& 0.05 \\
B & 100 	& 1 		& 0.315 	& 143.4 & ~ 0 & 0.14 	& 0.19 \\
C & 400 	& 0.75 		& 0.315     & 71.7 	& 0.3 & 0.07 	& 0.05 \\
D & 400 	& 1 		& 0.315 	& 143.4 & 0.1& 0.14 	& 0.19 \\
  \hline
\end{tabular}
\end{table*}

Tsiklauri (2011) simulated the propagation of inertial and kinetic Alfv\'en waves
through inhomogeneous collisionless plasma that mimics solar coronal loops. This study considered plasma over-densities (instead of cavities). There was a production of energetic electrons that can be a cause of the X-ray emissions from the solar
atmosphere. A recent study of parallel electron acceleration was conducted
with 3D simulations (Tsiklauri 2012). The study considers a low beta
plasma with left-hand polarized Alfv\'en waves of relatively high frequency
$\omega=0.3 \omega_{ci}$ propagating over sharp density gradients. (It is important to
notice that for a given wavelength,  the left-hand polarized mode has a lower
frequency than the right-hand polarized one, and that its frequency cannot be
larger than $\omega_{ci}$). This study confirms that when the gradients
typical size is $\sim c/\omega_{pe}$, parallel electric fields develop. They
cause parallel electron acceleration, and perpendicular ion heating. The
author suggests that this process is at work in the solar flare acceleration regions.

\begin{figure}   
\includegraphics[width=7cm]{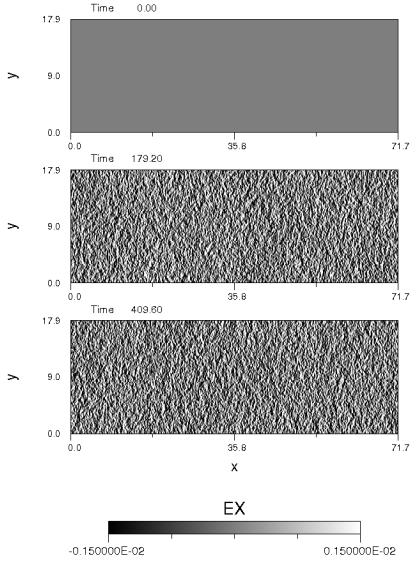}
\caption{Run A: snapshots from $t=0$ to $t=409.6$ of the parallel
   component of electric field  $E_x(x,y)$.}
\label{exunif}
\end{figure}

In the present paper, we present Particle In Cell (PIC) numerical simulations of the interaction of Alfv\'en waves
with an inhomogeneous plasma. The interplume structure is modeled by plasma with density gradients in the transverse direction to the background magnetic field. The plasma conditions are those expected in coronal hole region.
Our main goal is to explain the observed suprathermal electrons which are originate from coronal holes and particularly from interplume regions. The mechanism that we describe can contribute to the heating and the acceleration of solar fast wind.

The simulation model and parameters are exposed in section 2. The global electric field structure and the electron acceleration are shown in sections 3 and 4 respectively. Small scale coherent structures and their link with beam instabilities are exposed in section 5. Then we conclude.

\section{Simulation model and parameters}

\begin{figure}   
\includegraphics[width=7cm]{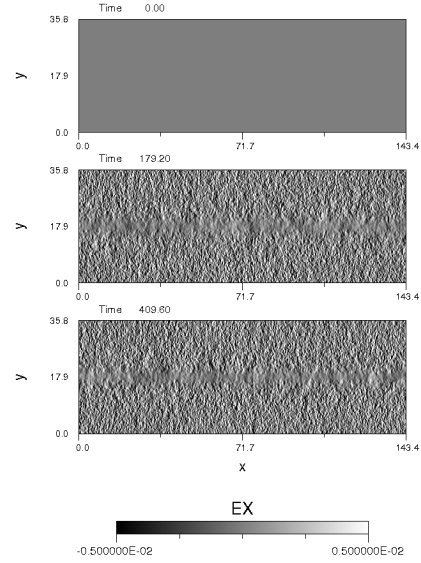}
\caption{Run B: snapshots from $t=0$ to $t=409.6$ of the parallel
   component of electric field  $E_x(x,y)$.}
\label{eyb}
\end{figure}

In the solar corona, the electron plasma frequency $\omega_{pe}$ is larger
than the electron cyclotron frequency $\omega_{ce}$. The ratio of the two
frequencies is given by the equation

\begin{equation}
\label{p1}
\frac{\omega_{pe}}{\omega_{ce}}\approx 3.2\times 10^{-3} \frac{n_0^{1/2}}{B_0}
\end{equation}

where the electron density $n_0$ is in cm$^{-3}$, and the magnetic field strength $B_0$ is in gauss (G).

In all the simulations, we assume a uniform background magnetic field $B_0$ in the parallel direction $x$, which is the solar radial direction.
The strength of the magnetic field is set to be $B_0=10$ G. The background electron density is given by $n_0 = 10^{8}$ cm$^{-3}$. 
The electronic temperature $T_e$ is assumed to  vary from $10^6$ to $10^8$ K,  which corresponds to $T_e=2$keV and $T_e=10$keV respectively (see Introduction for these plasma conditions).

\begin{figure*}   
\center
\includegraphics[width=7cm]{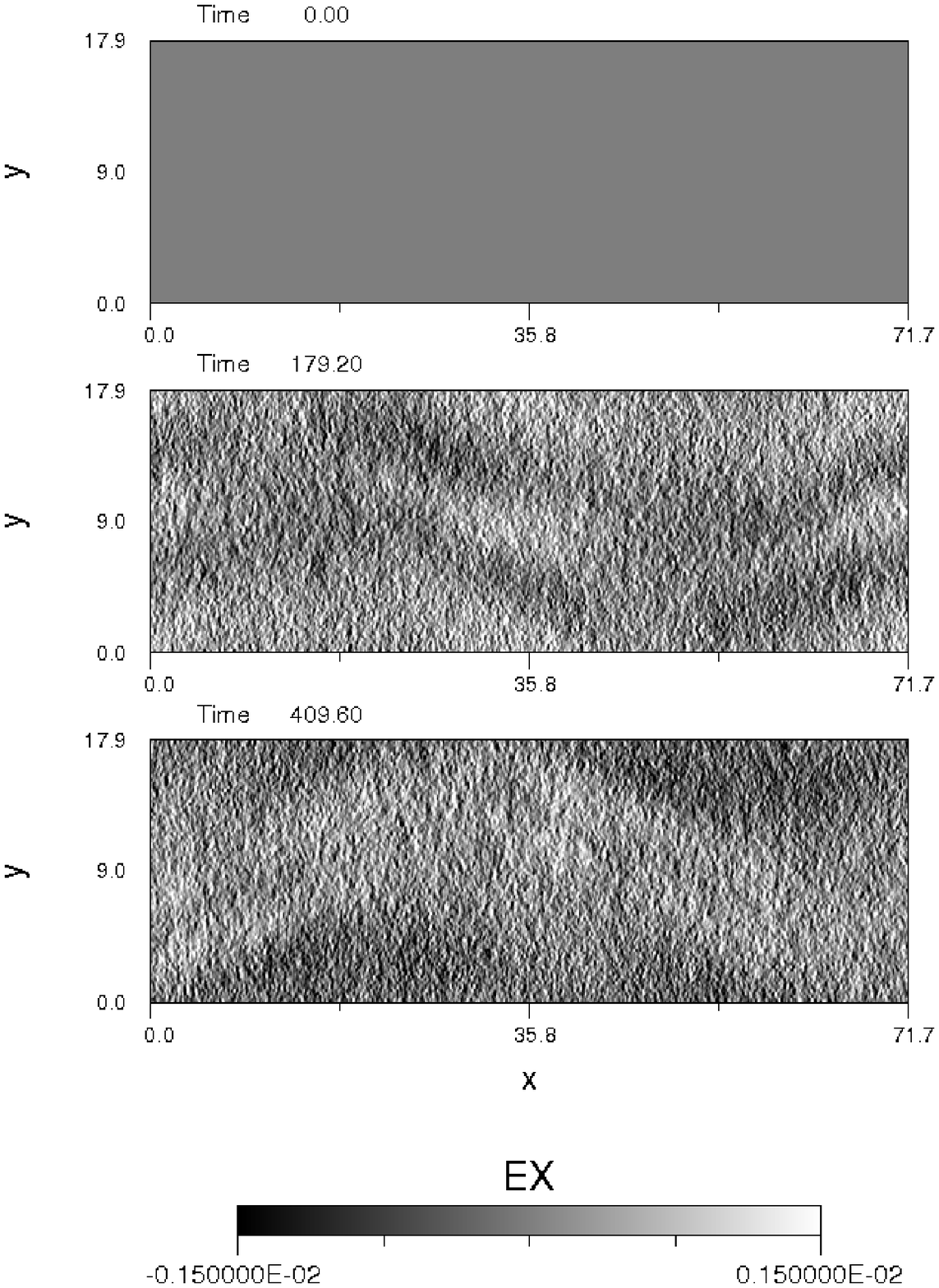}
\hspace{0.1\textwidth}
\includegraphics[width=7cm]{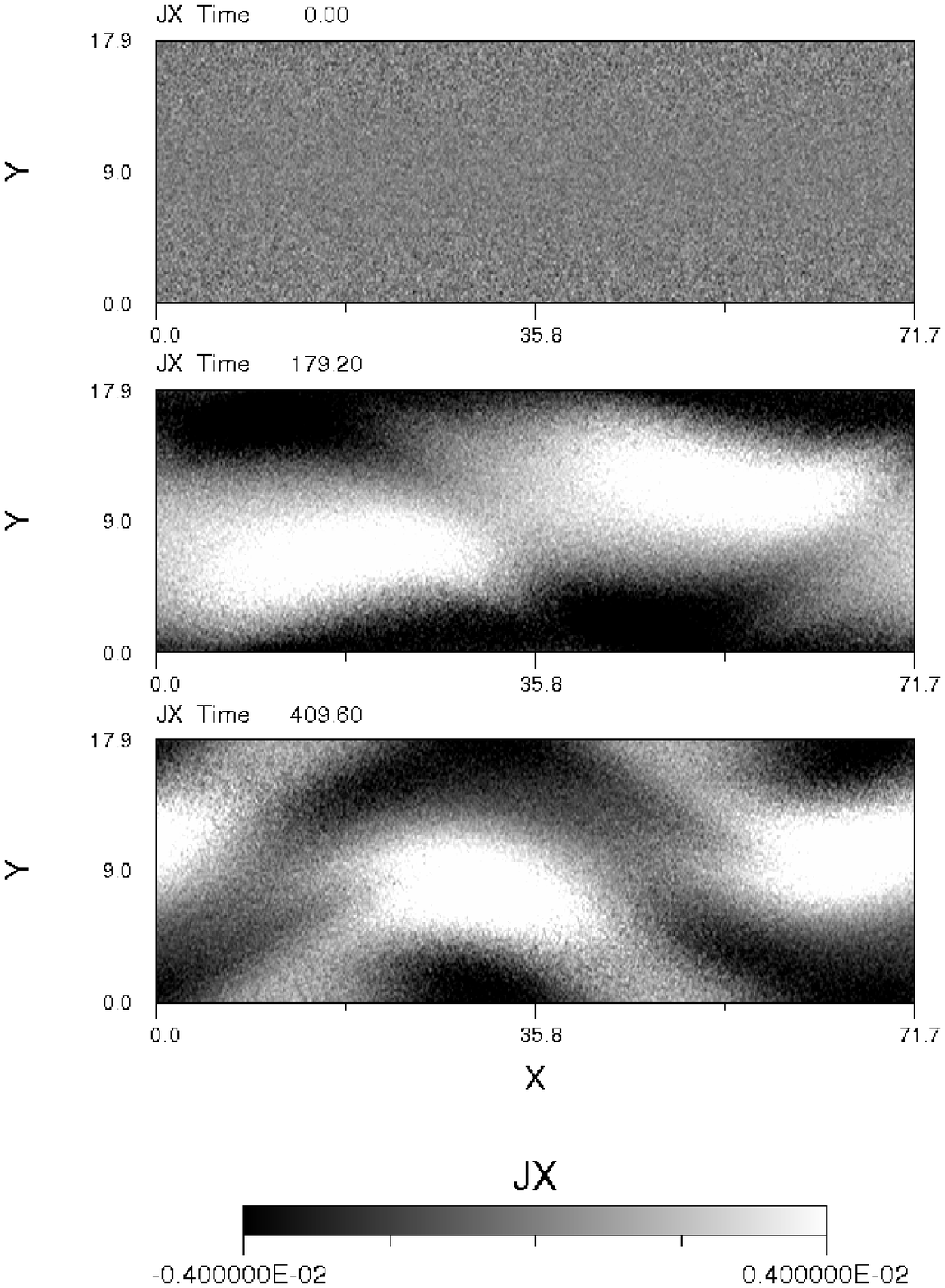}
\caption{Run C: snapshots from $t=0$ to $t=409.6$ of the parallel
   component of electric field  $E_x(x,y)$ in left panel, and the parallel current density associated to the electrons $J_x(x,y)$ in right panel.}
\label{exc}
\end{figure*}

When Alfv\'en waves propagate over a perpendicular density gradient with a
typical scale length that is below the MHD limits, inertial and
kinetic effects play a fundamental role in the creation of the parallel
electric field. Their behaviour is then very dependent on the parameter
$(m_i/m_e) \beta$, where  $\beta$ is the ratio of the kinetic pressure to the
magnetic pressure. With the plasma characteristics given above, $(m_i/m_e)
\beta <<1$. For instance, with $T \sim 2$ keV, $B_0=10$ G and $n_0=10^{8}$cm$^{-3}$,  and $\omega_{pe}/\omega_{ce}=3.17$ (Equation \ref{p1}), we find
$(m_i/m_e) \beta = (v_{Te}/c)^2(\omega_{pe}/\omega_{ce})^2 \sim 0.05$. We are
in the so-called inertial regime described by Goertz (1984). For $B_0=10$ G,
$n_0=10^{8}$cm$^{-3}$ and $ T \sim 10$ keV,  we find $(m_i/m_e) \beta = 0.2$ still corresponds to the inertial regime. It is therefore important that the plasma regime in the numerical simulations also corresponds to $\omega_{pe}/\omega_{ce} >1$ and $(m_i/m_e) \beta <<1$. 

However, when the temperature goes up to $10^8$ K we leave the inertial regime and reach the kinetic regime, but such temperatures are exceptional. The simulations 
presented in this paper are done in the inertial regime.  

We use an electromagnetic PIC code developped in Mottez, Adam \& Heron (1998) which takes into account the full electron and ion motion. The code is 2D in  space (($x,y$) coordinates) and 3D  in fields and velocity components. The boundary conditions are
periodic in $x$ and $y$ directions.

The physical variables used in the code are all dimensionless. The inverse of time and
frequency are normalized to electron plasma frequency $\omega_{p0}$  which
corresponds to the uniform background electron density $n_{0}$. 
The velocities are normalized to the speed of light $c$. The reference units
are the mass of the electron $m_e$ for the masses, $c /\omega_{p0}$ for
distances, $\omega_{p0}/c$ for the wave vectors, the electric
charge $e$ for charges, the electron density $n_0$ for charge densities,
$c\omega_{ce}/\omega_{p0}$ for the electric field. 
In what follows, in the absence of specific mention, all the figures and numerical values are expressed in dimensionless units.

We present a series of four numerical simulations whose parameters are summarised in Table
\ref{T-simple}.
The dimensionless background magnetic field $B_0$ is given by the frequency ratio
$\omega_{ce}/\omega_{pe}$. For $B_0=10$ G and  $n_0=10^{8}$ cm$^{-3}$, the  dimensionless
background magnetic field will be $B_0= 0.315$ (Equation \ref{p1}). 
The dimensionless thermal velocity $V_{Te}$ is
 0.07 or 0.14. The electron temperature $T_e$ and the ion temperature $T_i$ are equal. The simulation box size is $L_x \times L_y=1024 \Delta x \times 256 \Delta y$ where the cell size $\Delta x=\Delta y=\lambda_{De}=V_{Te}$ is one Debye length. For economy of computational resources, the ion to electron mass ratio is reduced to $m_i/m_e=100$.
The simulations are run over 2048 or 4096 time steps of duration $\Delta t=0.2$.

\begin{figure}   
\includegraphics[width=7cm]{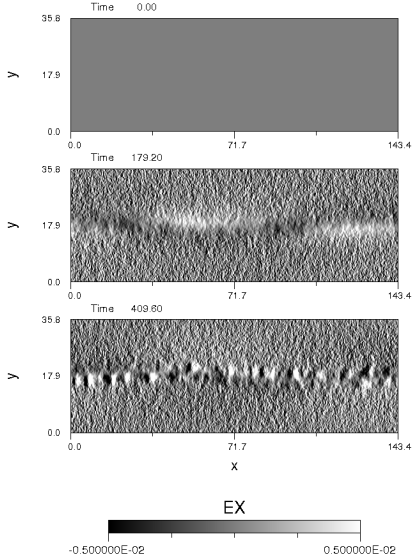}
\caption{Run D: snapshots from $t=0$ to $t=409.6$ of the parallel 
   component of electric field  $E_x(x,y)$.}
\label{exd}
\end{figure}

For the sake of simplification, we modelize a single interplume embedded between two plumes of identical density profile. This will give a cavity profile to the interplume density .

The interplume in Runs B, C and D is characterized by a plasma cavity of infinite length along the solar radial direction $x$  with a Gaussian
density profile in the transverse ($y$) direction

\begin{equation}
\frac{n(y)}{n_0} = 1-p [\exp\{-Y^2/R^2\}]
\label{nxycav}
\end{equation}

where $n(y)$ is the density on the interplume, $n_0$ is the density on the surrounding plumes axis . 

$Y=y-N_y/2$, where $N_y=256$ is the number of cells in $y$ direction and $p$ is the depth of the cavity $n_{min}/n_0=1-p$, and  
$R$ is the half of the cavity width. In all of these simulations, we have set $R=N_y/2$.

In simulations A, C, and D, we have initialised a right-hand polarized wave on
the Alfv\'en wave branch of the dispersion equation, which propagates from
left to right in the simulation domain. The details of this wave polarization
are given in Mottez (2008). In the initialisation, in accordance with the linear theory, there is no parallel electric field associated to this wave.  The
wave amplitude is controlled through the intensity of the sinusoidal perpendicular magnetic field $\delta B$. 
There is one wavelength in the box. The ratio 
of the wave frequency $\omega$ to ion cyclotron frequency is   $\omega/\omega_{ci} \approx
1.317$ for $V_{Te}=0.07$ (runs A and C); we are clearly not simulating a purely MHD Alfv\'en wave (it should correspond to $\omega/\omega_{ci} <<1$). We rather simulate an electron cyclotron wave, that is actually on the upper frequency part of the same dispersion relation branch as the MHD Alfv\'en wave. For $V_{Te}=0.14$ (runs B and D), the ratio  $\omega/\omega_{ci} \approx
0.54$. In all cases, our simulations relate to the high frequency part of the turbulent cascade of Alfv\'en waves, near the ion-cyclotron cut-off of the wave energy spectrum.

Nevertheless, in the paper, we speak of ``Alfv\'en waves'', even if we must keep in mind that we actually simulate a wave on the higher frequency part of the dispersion relation branch related to them.

Run C corresponds to a simulation with a
cavity of moderate amplitude ($p=0.75$), and runs B, D to a simulation
with a very deep amplitude ($p=1$) and a less intense
incident wave. 

In what follows, we focus mainly on electric field and on the velocity distribution of electrons.


\section{Electric fields}
\label{parallel}

Run A is a test of Alfv\'en wave propagation through a uniform plasma ($p=0$). Figure \ref{exunif} shows the parallel electric field $E_x(x,y)$ from $t=0$ to $t=409.6$ for the run A. During this propagation, no parallel electric field and no compression of the plasma is
observed in Figure \ref{exunif} in spite of quite large amplitude
    wave. Furthermore, the wave packet is not attenuated during the simulation and the wave vector remains parallel to the ambient magnetic field ($k_{\perp}=0$).

Opposite to the run A, the plasma in run B is inhomogeneous with the
presence of an interplume as a density cavity centred at $L_y/2$. However, no wave is set in the initial conditions.
Figure \ref{eyb} shows the component $E_x$ displayed in the $x-y$ plane. The snapshots are from $t=0$ to $t=409.6$. No parallel electric fields were observed during all the simulations. Nevertheless, small electrostatic oscillations were observed for the $E_y$ component in the transverse direction.

\begin{figure}   
\includegraphics[width=7cm]{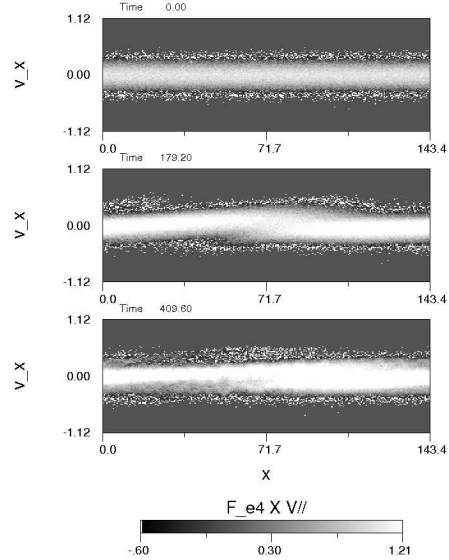}
\caption{Run D: snapshots from $t=0$ to $t=409.6$ of the electron distribution function $f_e(x,V_x)$ in logarithmic scale, at a fixed $y$ (Figure \ref{exd}).}
\label{fey}
\end{figure}

In runs B, C, D, we assume a pressure equilibrium condition for the cavity, as is the case of tangential discontinuities, which requires that the pressure inside the cavity is equal the pressure outside it. This condition implies a "correction" of the parallel background magnetic field as

\begin{equation}
\label{bx}
B_x=B_0 \left[1+\frac{\beta_0}{2}\left(1-\frac{n}{n_0}\right)\right].
\end{equation}

Furthermore, to insure an MHD stability of the channel, we
    assume that the density gradient scale is large in comparison with ion
    gyroradius for all simulations.
However, to construct a more stabilized tangential discontinuity, we need to apply plasma kinetic
equations with non trivial electron velocity distribution. This could allow sharp cavities that are authentic equilibrium solutions without electric field
  (Channell 1976; Mottez 2004; Sitnov et al. 2006; Liu, Liang \& Donovan
    2010; Kocharovsky, Kocharovsky \& Martyanov 2010),
  some of them being directly applied to plasma cavities (Mottez 2003). More
  recent computations show that sharp density gradients can also be associated
  to finite non-oscillating perpendicular 
electric fields (De Keyser, Echim \& Roth 2013).

Run C includes an interplume and  an Alfv\'en wave.  We note at first that the cavity is not destroyed by the Alfv\'en wave (not shown on a figure).
In Figure \ref{exc}, the parallel electric field $E_x(x,y)$ and the parallel current density associated to the electrons $J_x(x,y)$ are represented through
three different times.
At time 179, a significant parallel electric field is clearly visible on the edges of the cavity.
Along the $x$ axis, the parallel electric field ($E_x$) structure has the same wavelength as the initial Alfv\'en wave. 
These characteristics are the same as those seen in the numerical simulations 
in G\'enot et al. (2004) made in the context of the Earth auroral zone, and in
Tsiklauri et al. (2005) and Tsiklauri (2011) in the context of the solar corona.

As was shown previously (G\'enot, Louarn \& Mottez 2000; Tsiklauri 2011), the wave velocity is higher in the low density regions than outside the cavities. A distortion of the wave front results from this 
velocity shifts, creating a situation similar to those of an oblique Alfv\'en wave (G\'enot, Louarn \& Le Qu\'eau 1999; Tsiklauri 2007). This effect is also known as "phase-mixing" (e.g., Heyvaerts \& Priest 1983; Tsiklauri 2016a).
As can be seen in Table \ref{T-simple}, all the simulations are done in the inertial regime. 
The origin of the parallel electric field results from the bi-fluid theory of dispersive inertial Alfv\'en waves (Goertz 1984), that is relevant when $(m_i/m_e) \beta<<1$. 
In this context, two different opinions explain the origine of the parallel electric field $E_{||}$ and it dependance on the mass ratio $m_i/m_e$. G\'enot, Louarn \& Le Qu\'eau (1999) suggest that  with an oblique Alfv\'en wave, the polarization drift moves the ions in the transverse direction creating a space charge on the density gradient regions. A charge density is created, as well as the parallel electric field, and neutrality can only be restored by the longitudinal motion of electrons.  Then $E_{||}$ is proportional to the ion mass $m_i$ as the polarization drift equation. The second suggestation by Tsiklauri (2007) consider that the cause of $E_{||}$ generation is  electron and ion flow separation induced by the transverse density inhomogeneity, which is different from the electrostatic effect of charge separation. The author derive an analytic equation which explains the process of $E_{||}$  generation by ion-cyclotron waves. However, the obtained $E_{||}$ in this case decreases linearly as the inverse of $m_i/m_e$, i.e., proportional to $1/m_i$. 

Overall, phase-mixing in the transversely inhomogeneous plasma naturally creates $k_{\perp}$, which in turn leads to $E_{\parallel}$.

In simulation C, we have calculated the maximum separation ($\Delta y)_{max}$ between ions and electrons on the
density gradient (see G\'enot et al. 2004 for calculations). This value $\sim 11 \lambda_{De}$  is an estimate of the width of the acceleration
region in the transverse direction ($\lambda_{De}=0.07$).

\section{Electron acceleration}

Figure \ref{exd} shows the parallel electric field $E_x(x,y)$ for run D.

Figure \ref{fey} shows the electron distribution function  $f_e(x,V_x)$ for $N_y$/2 $<y<$ 5$N_y$/8, in logarithmic scale for Run D. In the snapshots at $t=179.2$ and $t=409.6$, we can distinguish clearly a beam of fast electrons 
escaping with velocities $V_x$ about 3 $V_{Te}$, which are larger than the incoming wave phase velocity ($V_A=0.0315$). Therefore,  a part of the accelerated electrons propagate temporarily in the front of the wave packet with a suprathermal velocity. In the case of Earth magnetosphere where the magnetic field is much higher, Mottez \& G\'enot (2011) found the same result but with a much higher velocity for the electrons beam (8$V_{Te}$). In the snapshot at $t=409.6$, the signature of instability is clearly seen in the form of a vortex structures associated to the electron beam. The electron holes associated to vortex structures trap the particles in their electric field and reduce the acceleration process. 

\begin{figure}    
\center
\includegraphics[width=5.9cm]{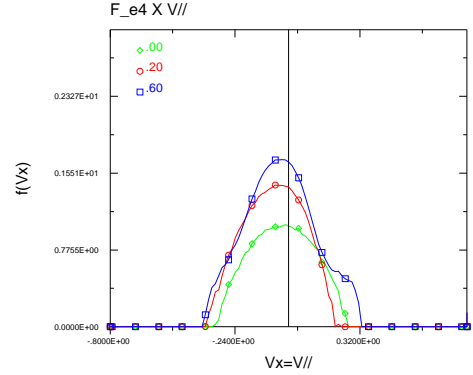} 
 \caption{Run D: a cut at $x=72$ of the electron distribution function $f_e(x,V_x)$ as a function of the parallel velocity $V_x$. The three color curves are for different values of $y$.}
\label{fex}
\end{figure}

Figure \ref{fex} shows a cut at $x=72$ ($N_x=512$) of the electron velocity distribution function for three different values of $y$. For the distribution function $y=0.6$ (blue curve), a second small peak is visible at $V_x\sim$ 0.26 $\approx$ 2$V_{Te}$ which corresponds to the accelerated beam electrons. 

This figure shows also a global shift of the distribution function in the
parallel direction, particularly for the red and the blue curves. This indicates that electrons gained energy from their interaction with the parallel electric field in gradient regions of the cavity. The velocity drift is $V_{drift}\approx -0.06$, which is smaller than the thermal velocity. This explain also why we do not observe a Buneman instability.  


\section{Electron beam plasma instability}
In run C, the typical size of the electric structure is those of the incident Alfv\'en wave ($\lambda=71.7$). In run D (Figure \ref{exd}), we observe smaller-scale
bipolar electrostatic structures along the interplume cavity. Their typical scale is about 30 to 45$\lambda_{De}$ which is twice or three times  smaller than the  structures found by G\'enot et al. (2004) in the context of the Earth auroral zone.

\begin{figure}      
\center
\includegraphics[width=7cm]{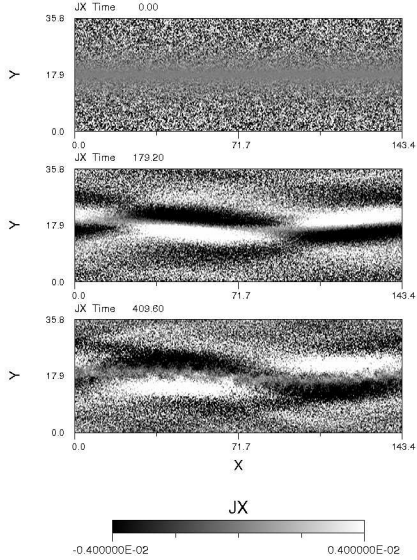} 
 \caption{Run D: snapshots from $t=0$ to $t=409.6$ of 
   the parallel current density associated to the electrons $J_x(x,y)$.}
\label{jxd}
\end{figure}

Figure \ref{jxd} shows the parallel current density associated to the electrons. The signature of a strong acceleration of electrons in the interplume cavity is clearly apparent from $t$ = 0 to $t$ = 179.2. However, at $t$ = 409.6, small transverse structures are visible. 

\begin{figure}    
\center
\includegraphics[width=7cm]{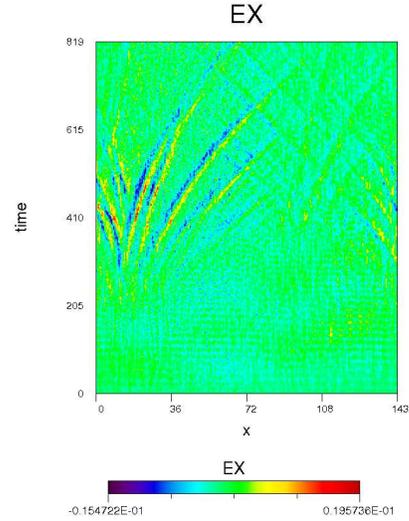} 
 \caption{Run D: the temporal variation of the component $E_x$ of the wave as
   a function of the  coordinate $x$ and time at $y=16.8$ ($N_y=120$).}
\label{Extinstb}
\end{figure}

Figure \ref{Extinstb} shows the temporal evolution of the parallel electric field in Run
D as a function of $x$,  at $y=16.8$ which is approximately the position of
the small structures observed in the lower part of the interplume cavity (Figure \ref{exd}). For $t<200$, we can
observe the evolution of the large scale parallel electric of the same size as the wavelength ($\lambda=143.4$). These fields
propagate from left to right sides as well as the Alfv\'en wave
propagates. After $t\sim 250$, the large scale electric field disappears behind small
bipolar structures of stronger amplitude.
We can infer that these structures observed also in Figures \ref{exd} and \ref{jxd}
are a consequence of the non-linear evolution of an electron beam plasma
instability. The accelerated electrons escape from acceleration regions and
create a beam which interacts with the plasma. The beam becomes unstable
causing a re-organization of the parallel electric field into small-scale quasi-linear structures.
After $t \sim 307$, these structures are accelerated into the right hand side direction for $x>15$, and they are accelerated into the left hand side direction for $x<15$.
For $t>341$, we can also observe very weak structures between $x=90$ and $x=108$ that propagates only in the right hand side direction (positive slopes). 

We have measured the velocity of these structures by measuring the positive and
negative slopes of the observed stripes:

For the largest slope, we have found $V_x=0.07 \approx V_{Te}/2 $ in left and $V_x=0.21 \approx
1.5V_{Te}$ in right. These last values are in
accordance with the velocity measured in Figure \ref{fex} for the beam propagating in the right hand side direction.

For the smallest positive slopes, we have found $V_x \approx 0.033$ for the stripes at $x=$ 15, and  $V_x \approx 0.0286$ for the weak stripes situated between $x=$ 90 and $x=$ 108 ($t > 334$).
This shows that another part of the electron population propagates around the phase velocity of the Alfv\'en wave ($V_A = 0.0315$), which indicates a signature of Landau damping due to the wave-particle interaction (see for example Tsiklauri 2016b) .

We can notice that the small scale structures present analogies with the coherent structures observed in the Earth auroral zone acceleration regions (Temerin 1982; Mottez et al. 1997; Ergun et al. 1998). Isolated bipolar electrostatic structures in the solar atmosphere and particularly in the solar wind were analysed in Mangeney et al. (1999) with the {\it{WIND}}/WAVES waveform analyser. It was found that these structures are isolated spikes of duration 0.3 to 1 millisecond. They consist of two main weak layers of approximately opposite charge with a spatial width $ \sim 25 \lambda_{De}$ close to the thickness of the structures observed in our simulation. Compared to the Earth magnetosphere environment, it seems that the background magnetic field strength is related to the size of these structures. The observations show also that these structures
propagate along the magnetic field line at a velocity smaller than the solar
wind velocity. However, the double layers  in the solar wind are seen in an
uniform plasma, while those that we have simulated are seen in the environment of a deep cavity.

We have to note that the limitation in the non-linear interactions in the space (2D) might favour the transfer of energy to parallel fluctuations, since the perpendicular direction is absent. In a more realistic 3D scenario, the conversion to the perpendicular electric field would be much more pronounced, reducing the acceleration effect in the parallel direction.


\section{Conclusion}

The detection of suprathermal electrons in the solar atmosphere remains a current problem. Different acceleration processes based on magnetic reconnection have been proposed. However, observations indicate that electrons are accelerated even in the absence of solar flares.
In this paper, we have presented 2.5 D particle-in-cell numerical simulations of Alfv\'en wave propagation through  an interplume region. The interplume structure
is modeled by a cavity density gradient perpendicular to the background magnetic field. We have started the simulations with conditions similar to those of the solar coronal holes. The aim is to contribute to the explanation of the observation of high energy electrons in solar corona through wave-particle interactions.

We have showed that an Alfv\'en wave propagating along a typical interplume in inertial regime is able to generate electric fields parallel to the ambient magnetic field  in the solar radial direction. The electric fields are localised in the density gradient regions with a typical size about the wavelength of the Alfv\'en wave. These electric fields can accelerate and heat electrons in the parallel direction.

In the case of strong amplitude cavity and less intense Alfv\'en wave, we have found that accelerated electrons can reach velocities about 2$V_{Te}$. They form beams and they generate a beam plasma instability whose non-linear evolution results in small scale electrostatic structures. 
 
These features can be identified as weak double layers. They present similarities with the spikes observed in the solar wind with {\it{WIND}}/WAVES waveform analyser. We note that despite the strong amplitude of the Alfv\'en wave propagating along it, the cavity is not destroyed. This indicates that interplumes are stable structures that could survive during the Alfv\'en waves propagation.

Modeling the acceleration source in interplume regions is a real challenge regarding the lack of high spatial resolution by solar observatories. In our simulation, the largest transversal width of the interplume cavity is of order of 256 $\lambda_D$, which corresponds to 388 m. The today highest spatial resolution measured with {\it{TRACE}} is about 370 km, which makes the observation of such small scale structures impossible.  However, many characteristics of interplume structures remain poorly understood like their cross-sections and their life. Possibility of fractal nature of plumes was invoked by some authors (e.g., Llebaria et al. 2002; Milovanov \& Zelenyi 1994; Boursier \& Llebaria 2008), which implies the possibility of existence of interplume microstructures. The future high spatial resolution measurements by Solar Probe Plus and Solar Orbiter space missions may resolve these fine structures, and provide further details about the heating and the acceleration of particles in interplume regions.


\acknowledgements

The authors thank the anonymous referee for constructive comments and suggestions that improve the quality of the paper.


\appendix

\end{document}